\documentclass{IEEEtran}

\usepackage[pdftex]{graphicx}
\usepackage{multirow}
\usepackage{tabularx}
\usepackage{todonotes}
\usepackage{booktabs}
\usepackage{hyperref}

\ifCLASSINFOpdf
\else
\fi
\begin{document}
%
\title{High Throughput 2D Spatial Image Filters on FPGAs}


\author{\IEEEauthorblockN{Abdullah~Al-Dujaili}
\IEEEauthorblockA{\\School of Computer Science and Engineering\\
              Nanyang Technological University, Singapore\\
Email: aldujail001@e.ntu.edu.sg\\}
\and
\IEEEauthorblockN{Suhaib~A.~Fahmy}
\IEEEauthorblockA{\\School of Engineering\\
              University of Warwick, United Kingdom\\
Email: s.fahmy@warwick.ac.uk}
}


%


\maketitle

\begin{abstract}
FPGAs are well established in the signal processing domain, where their fine-grained programmable nature allows the inherent parallelism in these applications to be exploited for enhanced performance.
As architectures have evolved, FPGA vendors have added more heterogeneous resources to allow often-used functions to be implemented with higher performance, at lower power and using less area.
DSP blocks, for example, have evolved from basic multipliers to support the multiply-accumulate operations that are the core of many signal processing tasks.
While more features were added to DSP blocks, their structure and connectivity has been optimised primarily for one-dimensional signal processing.
Basic operations in image processing are similar, but performed in a two-dimensional structure, and hence, many of the optimisations in newer DSP blocks are not exploited when mapping image processing algorithms to them.
We present a detailed study of two-dimensional spatial filter implementation on FPGAs, showing how to maximise performance through exploitation of DSP block capabilities, while also presenting a lean border pixel management policy.
\end{abstract}


%
\IEEEpeerreviewmaketitle

\section{Introduction}
\label{intro}
Image and video processing generally involve intensive computations on large amounts of data.
Consider a streaming colour video signal at a spatial resolution of 1920$\times$1080 pixels and a rate of 60 frames per second -- that represents a throughput of over 124 million pixels per second.
With many operations required per pixel in a typical vision flow, this represents a computational requirement of many GOPS if real-time processing is required.
Hence, parallelism must be exploited for such systems to be feasible in real-time~\cite{crookes,weems,smarter_vision_requirement_video}.
Spatial filtering is a fundamental approach used in the lower stages of many vision applications, and hence, if it is inefficient, it can hamper the higher layers in such algorithms.
The fidelity of video data is also rising significantly, with 4K video now within reach of a mainstream audience, quadrupling the computational requirements compared to 1080p.
All this points to the fact that optimised spatial filtering is an important part of real-time vision systems.

Implementing high-speed image-processing systems on Field Programmable Gate Arrays (FPGAs) has been a very active field.
This is mainly due to the ability to leverage bit-level, pixel-level, neighbourhood-level, and task-level parallelism to accelerate computation.
Moreover, FPGAs are reconfigurable, allowing for the flexibility often desired in vision systems.
This coupling of very high speed processing, parallelism, and flexibility is what makes FPGAs a platform of choice in real-time vision systems~\cite{Crookes_intro,suhaib_median_filter}.

As processing moves up the vision pyramid, from pixel-level operations, to more advanced algorithms on less data, software implementations can be more attractive, due to ease of programming, and irregular data access.
Hence, the ideal real-time vision platform would couple high throughput, highly parallel low-level pixel operations implemented in custom hardware, with a tightly coupled processor to take care of higher level operations.
New hybrid FPGAs like the Xilinx Zynq provide a very capable embedded processor with a flexible reconfigurable fabric on the same silicon, with high throughput connectivity between them.
This represents an ideal platform for embedded implementation of the full computer vision stack including higher level software with low-level hardware.
Within this context, we explore generalised blocks for low-level operations that maximise throughput to support advanced vision algorithms on high bandwidth video streams.

One of the most frequently used low-level operations is 2D linear spatial filtering (convolution), in which a pixel in the output image is determined by applying a spatial arrangement of coefficients to the neighbourhood of the same pixel in the source image.
Selection of suitable coefficients allows this same structure to be used for many of the typical low-level processing operations in a vision system, such as noise removal, image sharpening, blurring/smoothing, and feature extraction.
As spatial filtering is typically applied on a streaming image, and requires no intermediate storage of complete frames, it is important to optimise this operation so it does not become the bottleneck in full vision system implementation.
Furthermore, building a generalised structure where coefficients can be changed at runtime allows the same hardware to be used for multiple tasks.
Much of the previous work on FPGA-based spatial filters attempts to minimise area and improve throughput by fixing the coefficients and optimising the filter structure accordingly.
This does not fit well within the context of a smart vision system where the coefficients would be adapted based on information from the higher layers, and multiple different iterations may be reuired.
We show that using modern DSP blocks it is possible to have a general filter that also offers very high throughput.

\begin{figure}[h!]
\centering
\includegraphics [width=3.5in] {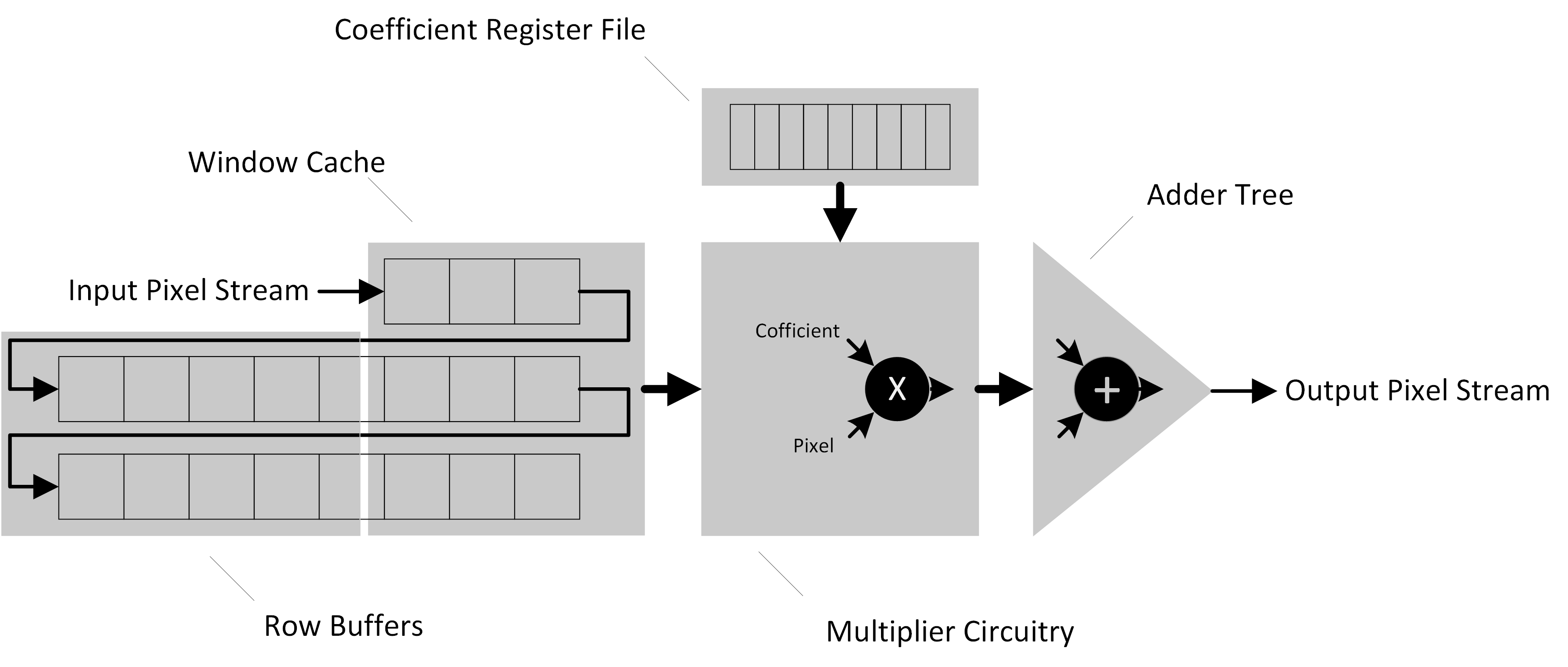}
\caption{Image filter block diagram.}
\label{fig_blk}
\end{figure}

A filter implementation is typically composed of four hardware blocks, as shown in Fig.~\ref{fig_blk}.
A stream of pixels from the source image is received, one per clock cycle, in raster scan order.
In order to determine an output pixel we require all pixels in the corresponding input pixel's neighbourhood to be available.
For a window size of $w$ (where $w$ is an odd number), this means we must buffer pixels from rows prior and subsequent to the desired output pixel's row number, with pixels from $w$ rows being available. This requires a row-buffer, with the minimal requirement of $w-1$ rows to be stored (since the older pixels from the first row can be discarded).
This avoids the need for full-image buffering, which for large frame sizes can become a constraining factor.

\begin{figure}[h]
\centering
\includegraphics[width=2.75in]{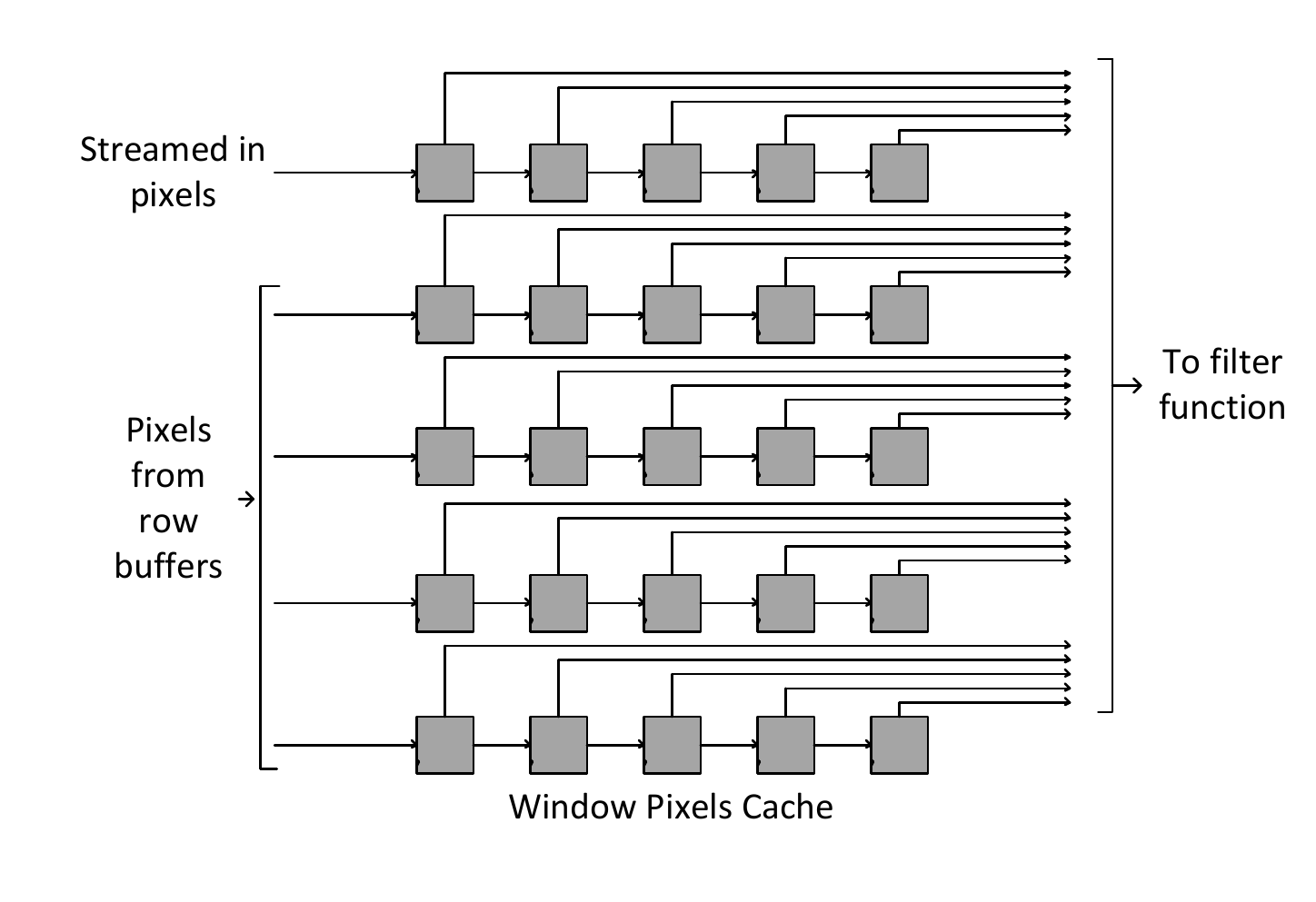}
\caption{A typical 5$\times$5 window pixel cache.}
\label{fig_ca}
\end{figure}

The pixels used to determine the output pixel represent a $w\times w$ window around and including the corresponding input pixel, shown as the window pixel cache.
Each of these pixels is multiplied in parallel by a corresponding coefficient, determined by the desired function, with this computation happening in the filter function block.
The coefficient file provides the filter function with the coefficients to use for convolution.
In this context, the coefficient file can be updated from the higher layers of the vision algorithm to alter the effect of the filter.
The control unit, implemented as a state machine, controls filter operation, including priming, flushing, activation, and deactivation.
At each clock cycle, we can compute one output pixel result, and this structure allows us to preserve the streaming data movement, eliminating the need for a full frame to be stored, and maintaining the input pixel rate.
Fig.~\ref{fig_ca} shows the arrangement of pixels in the buffers as supplied to the filter function.

The core computation in linear filtering involves multiplication and addition.
A number of approaches to efficient design of filters have been proposed.
Generally, filter designs can be split into two categories \cite{Multiplierless_categories}: multiplier-based filters, which implement multiplications explicitly through a set of multipliers as in  \cite{high_per_multipliered,pipe_multipliered,reducing_complexity_multipliered}, and multiplierless filters which avoid using expensive multipliers through various arithmetic transformations and representations as in \cite{CSD_mutliplierless,DA_mutliplier_less,add_shift_multiplier_less,efficient_fir_multiplierless}.

In modern FPGAs, multiplications and subsequent accumulations can be efficiently implemented using DSP blocks.
While DSP blocks were a scarce resource in the past, even small FPGAs today can include sufficient DSP blocks to implement large spatial filters.
Recent developments have seen these DSP blocks become even more sophisticated, allowing more types of computation to leverage them.
The DSP48E1, found in all 7-series devices from Xilinx, for instance, supports multiply, multiply-accumulate (MAC), multiply-add, and three-input-add functions, among others.
The architecture also supports cascading of multiple DSP48E1 slices to allow for wider computations and complex arithmetic without the use of general FPGA logic.
Considered design allows use of DSP blocks at near maximum performance (up to 550 MHz) or with power efficiency in mind~\cite{dspslice}.
Unfortunately, much of the focus of existing approaches, and even of the design of the DSP block, has been tailored for one-dimensional FIR filters.
An example is the cascade wires that allow subsequent DSP blocks to be chained without signals being routed through the logic fabric.
These are hard-wired in a manner that works for one-dimensional but not for two-dimensional filters.
As a result of such optimisations, image filters are unable to take advantage of a number of these features.

\begin{figure}[h]
\centering
\includegraphics[width=2.75in]{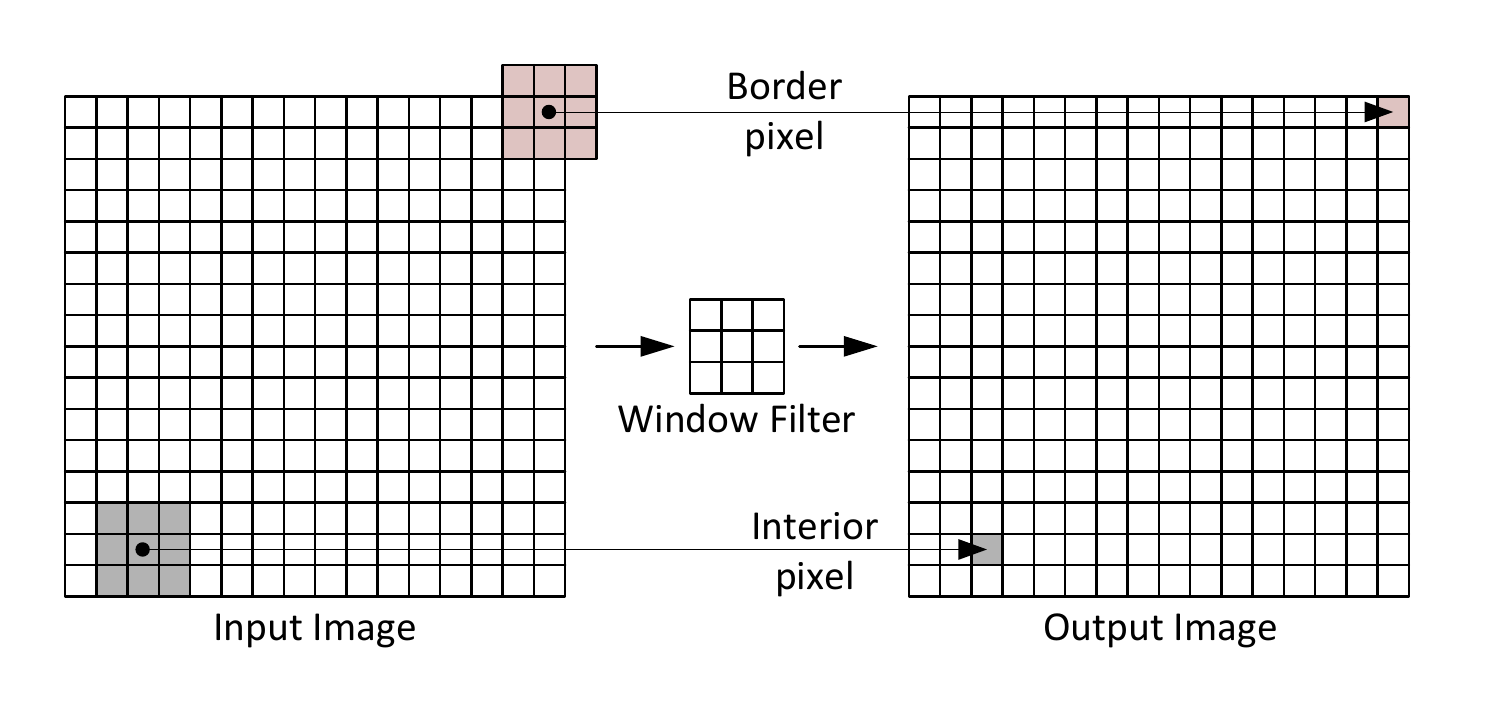}
\caption{Filtering for interior and border pixels.}
\label{fig_conv}
\end{figure}

In addition to the filter computation, window boundary handling is another important issue to be considered in any two-dimensional filter architecture.
As shown in as Fig. \ref{fig_conv}, each pixel in the output image is produced as the window goes over the corresponding pixel in the input image.
For interior pixels, each neighbourhood pixel is surrounded by valid pixels within the source image boundary.
For border pixels, however, some neighbourhood pixels may be outside of the image boundary.
Such cases must be considered during filter computation, requiring additional circuitry that imposes both area and performance overheads.

This paper explores the use of DSP blocks within the filter function, including the adder tree.
We explore the benefits and restrictions of traditional direct form and transpose form implementations, and determine how to extract maximum performance from the DSP48E1 while managing border pixels to minimize area and performance overheads.
We demonstrate overall filter throughput that is close to the theoretical maximum, while maintaining flexibility in terms of coefficients.

\section{Filter Function Implementation}

As mentioned earlier, filters can be implemented with or without multipliers.
Since DSP blocks are abundantly available in modern FPGAs, we explore multiplier-based implementations.
Note that for some spatial filters, where the coefficients are all zero or one, or other powers of two, such an architecture would not be necessary, since each multiplication could be replaced by a shift.
However, other applications require fractional values of higher precision to be applied over the filter window, and in such cases, multipliers are feasible.
Other multiplierless methods also force a fixed set of coefficients, meaning the filter block is single-purpose.
By using multipliers, our implementation is coefficient-independent and hence general-purpose.
Current DSP blocks are equipped with multipliers, adders, and cascaded connections~\cite{dspslice} which allow for one-dimensional transpose form filters to be implemented with no external logic.
Transpose form moves the adders to before the multiplication operation, and removes the need for a separate adder tree.

However, in image filtering, the input samples' timing and movement differ from those in one-dimensional filters.
Input samples are taken across multiple rows depending on the width of the filter window.
In streaming systems, this demands an appropriate delay of the results obtained from processing each row.
Therefore, the advantage of cascaded paths cannot be fully exploited.
On the other hand, implementing a direct form filter can increase  logic resources and power, since a separate adder tree must be implemented. The depth of the adder tree scales by $log_{2}$ of the window size $w$.

We have implemented both types of filters to explore how they map to DSP48E1 blocks in modern FPGAs:
\begin{itemize}
\item \textbf{Direct Form}: a set of multipliers compute pixel-wise products, while a separate adder tree adds these products to obtain the output pixel value. Multiplications are implemented using DSP blocks. Additions are implemented in three different ways, as shown in Fig.~\ref{fig_addr_blk}. As a result, we have three layouts for the adder tree:
	  \begin{enumerate}
	  \item \textbf{DSP} layout: using DSP blocks only through custom instantiation, each implementing a wide adder. DSP blocks come with various features such as three-input adder/subtracter. However, most of these features support only asymmetrical (different bit width) inputs or require aggregating DSP blocks via the internal bus~\cite{dspslice}, which limits design flexibility and scalability. Since the DSP block implements its additions directly in silicon, we use this as a baseline for maximum performance. In this arrangement, each DSP block is configured as a two-input adder using the post-adder block only. This is done by manually instantiating the DSP48E1 primitive and enabling only the post-adder path.
	  \item \textbf{LOG} layout: we implement the adder tree adders using the FPGA logic fabric, allowing for fine-grained control, and saving DSP blocks for other uses. This layout also balances the DSP block and logic utilisation better. The adder tree is described in general Verilog and the tools map this to LUTs (and the required registers for pipelining).
	  \item \textbf{DSPCOMP} layout: similar to \textbf{DSP} as it uses DSP blocks to perform additions, but with a logic-based compression step, that helps minimise  adder tree depth and the number of DSP blocks used. Here, the 6:3 compressor takes 6 operands and computes 3 partial sums, which are then summed through two DSP-based adders as for the DSP method.
	  \end{enumerate}

The adders in each of the three implementations are characterized in Table~\ref{table:addr_blk_spec}. The latency for each adder can be intuitively estimated from Fig.~\ref{fig_addr_blk}. As our goal is to achieve a high throughput, we heavily pipeline the basic blocks.
As a result, DSP blocks have a latency of 4 cycles for addition, fabric-based adders need 1 cycle, and compression logic takes 2 cycles to produce the compressed signals.
For \textbf{DSP}, the number of blocks required depends on data width as a single DSP block can be configured to be a 48-bit single, 24-bit dual, or 12-bit quad adder block in SIMD mode. Therefore, provided that the operand width is less than or equal 24 bits, fewer DSP blocks are required.

\begin{figure}[h]
  \centering
  \includegraphics[width=\linewidth,scale=0.5]{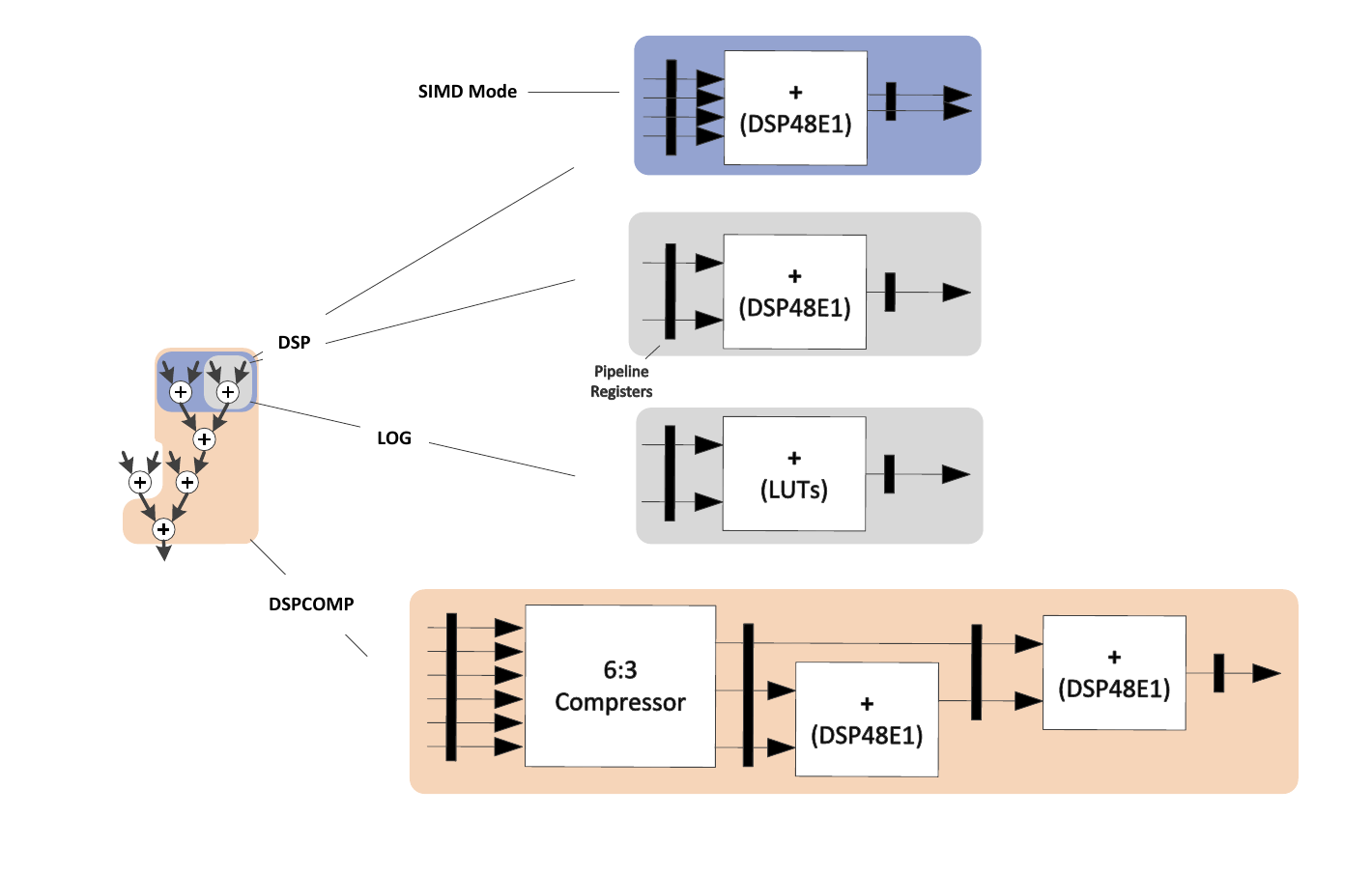}
  \caption{Adders for different adder tree layouts. From top to bottom: \textbf{DSP}, \textbf{LOG}, and \textbf{DSPCOMP}.}
  \label{fig_addr_blk}
  \end{figure}

   \begin{table}[h!]
    \small
   \caption{Adder specifications for adder tree layouts.}
    \label{table:addr_blk_spec}
  \begin{minipage}[b]{1.0\linewidth}\centering
  \renewcommand{\arraystretch}{1.2}
    \centering
  \begin{center}
  \begin{tabular*}{1.01\linewidth}{p{1.0in}ccc}
  \toprule
  & \multicolumn{3}{c}{Implementation} \\
  \cmidrule{2-4}
   & DSP & LOG & DSPCOMP \\

   & \multicolumn{1}{l}{} &  \\
  \midrule
  Number of Inputs & 2 & 2 & 6 \\
  \midrule
   \multirow{2}{*}{Basic Units} & \multirow{2}{*} {1 DSP48E1} & \multirow{2}{*} { LUTs} & 2 DSP48E1s \\
  & & & LUTs \\
  \midrule
  Latency  & 4 & 1 & 10\\
  \midrule
  Number of adders for $w=7$  & 48/36* & 48 & 10\\
  \midrule
  Number of stages for $w=7$   & 5 & 5 & 3\\
  \bottomrule
  \multicolumn{4}{l}{*without/with SIMD mode}
  \end{tabular*}
  \end{center}
   \end{minipage}
   \end{table}

\item \textbf{Transposed Form}: By reversing the signal flow and arranging the filter building blocks accordingly, the individual pixel products can be added as soon as they are computed. Hence, a multiply-accumulate operation is required, matching the DSP block architecture. A single multiplication and a single addition are packed into a single DSP block.
In a one-dimensional implementation, transpose form completely eliminates the need for external logic, using only DSP blocks.
Since the connections are different for a two-dimensional filter, some logic is still consumed, but there are savings due to the lack of need for an adder tree.
\end{itemize}

Table~\ref{table:dsp_block_general} shows how many DSP blocks each implementation consumes for a filter window of $w\times w$ pixels.
The first term ($w^2$) represents the multipliers used for the individual pixel products.
Other terms show the multipliers used in the tree adder.
In \textit{Direct DSP}, it is possible to make use of the dual 24-bit SIMD mode (two 2-input adders) to pack two addition operations within a DSP block as long as there is no overflow hazard. In our design, this mode was used for the first stage of the tree adder.
In \textit{Direct LOG}, the tree adder does not contribute to the consumption of the DSP blocks.
In \textit{Direct DSPCOMP}, the compressors allow us to pack a 6-input adder into \textbf{two} DSP blocks compared to \textbf{five} in \textit{Direct DSP}.
These expressions hold as long as the overflow condition is satisfied. DSP block usage for a $7 \times 7$ window is also shown in Table~\ref{table:dsp_block_general}.

 \begin{table}[h!]
 \small
 \hfill{}
 \caption{DSP Blocks usage for different filter function implementations for window of $w\times w$ pixels.}
  \label{table:dsp_block_general}
\begin{minipage}[b]{1.0\linewidth}\centering
\renewcommand{\arraystretch}{1.2}
  \centering
\begin{center}
\begin{tabularx}{\linewidth}{p{0.1in}XcXc}
\toprule
\multirow{3}{*}  &  & \multicolumn{2}{@{}c@{}}{DSP Blocks Usage} & \multirow{3}{*} {\begin{tabular}{@{}c@{}}DSP blocks for\\[-2pt] $w=7$ \end{tabular}} \\
\cline{3-4}
& &\multirow{2}{*}{\begin{tabular}{@{}c@{}}Multipliers\\[-2pt] Circuit \end{tabular}}  & Adder Tree & \\ 
\\
\midrule Direct & \\
\cline{2-5}
& DSP & $w^2$ &$\frac{w^2-1}{4}+\frac{w^2-1}{2}$ & {\begin{tabular}{@{}c@{}} 85 \end{tabular}} \\
\cline{2-5}  		& LOG & $w^2$ & - & {\begin{tabular}{@{}c@{}} 49 \end{tabular}}\\
\cline{2-5} & DSPCOMP & $w^2$& $\lceil \frac{w^2-1}{5}\rceil$ & {\begin{tabular}{@{}c@{}} 69 \end{tabular}} \\
\midrule Transposed &  & \multicolumn{2}{@{}c@{}}{$w^2$} & {\begin{tabular}{@{}c@{}} 49 \end{tabular}} \\
\bottomrule
\end{tabularx}
 \hfill{}
 \end{center}
 \end{minipage}
 \end{table}

Table \ref{table:latency_general} shows the latency in clock cycles as a function of window size $w$, image width $IW$, multiplier latency $M_L$ (3 cycles for our implementation), and adder latency $A_L$ (with values shown in \ref{table:addr_blk_spec}).

We assume that border pixels are taken care of in the \textit{Direct Form} and are discarded in the \textit{Transposed Form}. In other words, For an $H\times W$-image, \textit{Direct Form} outputs an $H\times W$ image, whereas \textit{Transposed Form} outputs an $(H-\frac{w-1}{2})\times (W-\frac{w-1}{2})$ image. This adds an additional latency of $\frac{w-1}{2}\times IW$ cycles to produce the first output pixel in  the \textit{Transposed Form} compared to the \textit{Direct Form}.

In \textit{Direct Form} implementations, addition and multiplication operations are performed separately. On the other hand, in \textit{Transposed Form} implementation, both computations are packed and we denote their latency altogether as ($C_L=3$). We also show the latency value for an image of $100 \times 100$ and $w=7$. 


 \begin{table}[h]
 \small
 \hfill{}
 \caption{Latency for different filter function implementations.}
  \label{table:latency_general}
  \begin{minipage}[b]{1.0\linewidth}\centering
  \renewcommand{\arraystretch}{1.2}
    \centering
  \begin{center}
\begin{tabularx}{\linewidth}{p{0.2in}p{0.5in}p{1.15in}c}
\toprule
\multirow{2}{*}{Implementation} &  & \multirow{2}{*}{ Latency (cycles)} & \multirow{2}{*}{ \begin{tabular} {@{}c@{}} Latency for $w=7$
\\, $IW=100$
\end{tabular}} \\
\\
\midrule Direct \\ \cline{2-4}
	& DSP &
	$\frac{w-1}{2}\times IW+\frac{w+1}{2}+ M_L + A_L \times \log_2(w^2)
	$ & 331\\
	\cline{2-4}

	& LOG &
	$\frac{w-1}{2}\times IW+\frac{w+1}{2}+M_L+A_L \times \log_2(w^2)$ & 313 \\
	\cline{2-4}

	& DSPCOMP &
	$\frac{w-1}{2}\times IW+\frac{w+1}{2}+M_L+A_L \times \log_5{w^2}$ & 337 \\
	\midrule

	Transposed &  &
	$(w-1)\times IW+w+C_L$ & 610 \\
\bottomrule
\end{tabularx}
 \hfill{}
 \end{center}
 \end{minipage}
 \end{table}

\section{Border Management Techniques}
\label{sec:border management technique}

While implementations of image filters on FPGA are many, few papers have tackled border management.
If border management is ignored, the output image will be smaller than the input image, since pixels with incomplete input neighbourhoods will be invalid, and should be discarded.
In one-dimensional filters, these initial invalid outputs are ignored since they only happen at the very beginning of a signal.
In image processing, however, this happens at the borders of every frame in a video, and hence must be considered.
In \cite{bailey_border}, the author reviews different border handling methods for FPGA-based 2D signal (image) filters and introduces a novel border management scheme with overlapped priming and flushing.
The method uses temporary pixel buffers (registers) and multiplexers to reduce the time overhead in handling borders pixels.
Another approach in \cite{Benkrid} considers symmetric extension for 1D signal border management through exploiting the SRL16 shift register primitives in Xilinx FPGAs to skew data.
However, this technique does not suit DSP block implementations as it inserts shift registers between the multiplication and addition operations, preventing efficient mapping to DSP blocks.

There are various general approaches to dealing with image borders.
Table \ref{IMAGE_BORDER_HANDLING_TECHNIQUES} lists most of these methods along with their advantages and disadvantages.
These methods are further detailed in \cite{bailey_book}.

 \begin{table}[h!]
 \renewcommand{\arraystretch}{1}
 \caption{IMAGE BORDER HANDLING TECHNIQUES}
 \label{IMAGE_BORDER_HANDLING_TECHNIQUES}
 \begin{minipage}[b]{1.0\linewidth}\centering
 \renewcommand{\arraystretch}{1.2}
 \begin{center}
\begin{tabularx}{\linewidth}{XXX}
\toprule
Technique & Advantages & Disadvantages
\\
\midrule

Border Neglecting
&
No additional control logic %
&
Reduced image size which can be problematic for small images or when cascading filters. %

\\
\midrule

Wrapping
&
Small control logic, Same image size
&
Possible discontinuity and artefacts
\\
\midrule

Function Change
&
Same image size
&
Complex control logic, difficult to generalize to all filters
\\
\midrule

Constant Extension
&
Same image size
&
Discontinuity, artefacts, additional control logic
\\
\midrule

Border Duplication
&
Same image size
&
Discontinuity, artefacts, additional control logic
\\
\midrule

Mirroring with/without Duplication
&
Same image size
&
Additional control logic
\\
\bottomrule
\end{tabularx}
\end{center}
\end{minipage}
\end{table}

\begin{figure}[h]
\centering
\includegraphics[width=3.55in]{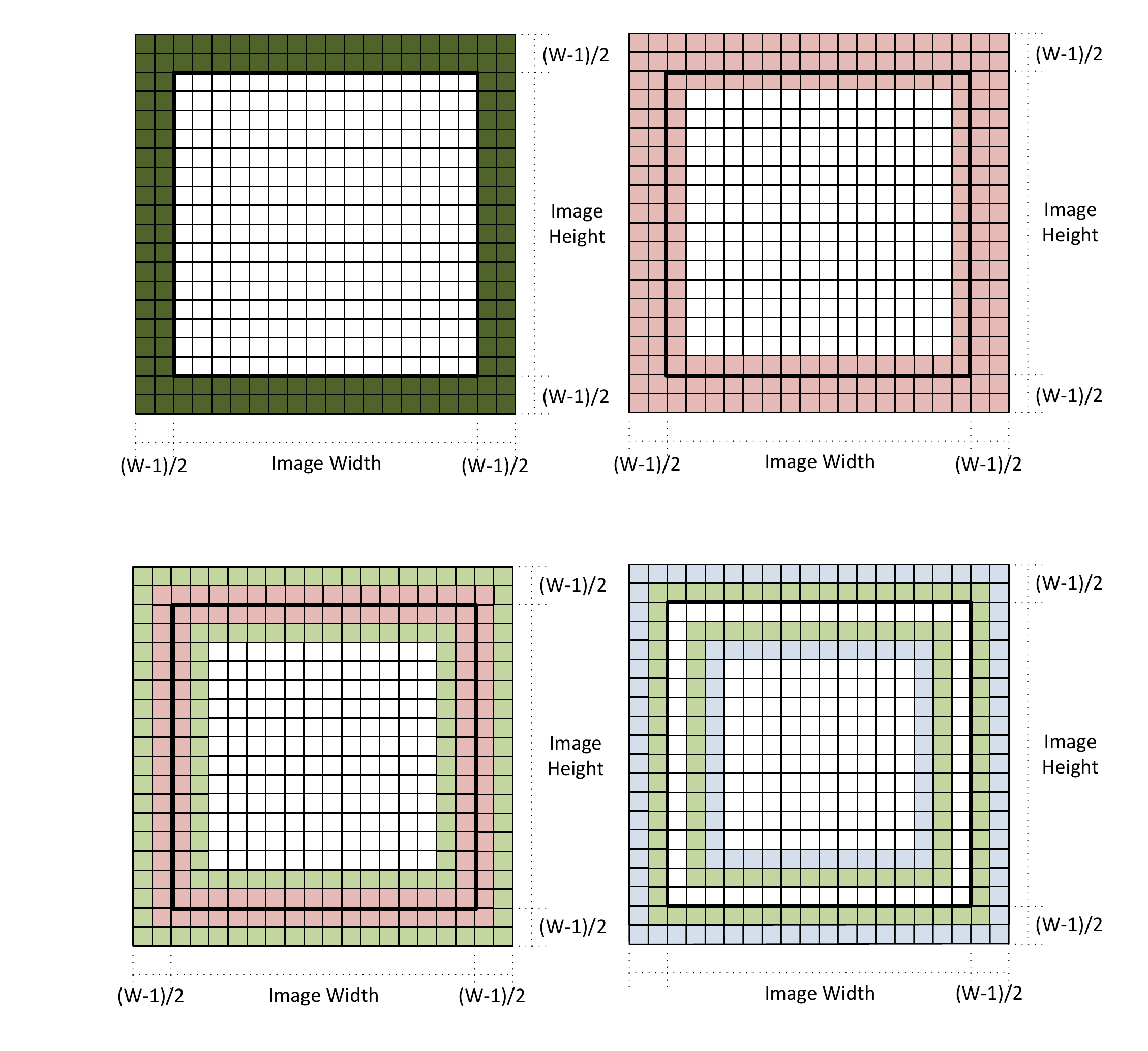}
\caption{Border management techniques (top left: constant extension, top right: border extension, bottom left: mirroring with duplication, bottom right: mirroring without duplication).}
\label{fig_tech}
\end{figure}

The last three techniques depicted in Fig. \ref{fig_tech} are generally more widely accepted, as they offer improved results.
Mirroring, for instance, is used in \cite{Verma}, \cite{Kurak} and \cite{McDon}.
We consider border extension using both mirroring techniques in this paper.

 \begin{table}[t!]

 \renewcommand{\arraystretch}{1.3}
 \caption{Image border handling technique implementations in hardware for a window size of $w$.}
 \label{table3}
 \centering
\begin{tabularx}{\linewidth}{XXX}
\toprule
Scheme
&
Advantages
&
Disadvantages
\\
\midrule
Direct Window Input
&
No modifications to pixel cache
&
Complex address generation logic, stalling input stream when processing border pixels by ($w-1$) between rows and frames for priming and flushing
\\
\midrule
Cached Priming
&
No complex address generation logic
&
Stalling input stream when processing border pixels by $(w-1)/2$ between rows and frames for flushing, requires extra multiplexers
\\
\midrule
Naive Overlapped Priming \& Flushing
&
No stalling, no complex address generation logic
&
Extra multiplexers, extra temporary pixel buffers within pixel cache, and extra temporary row buffers
\\
\midrule
Overlapped Priming \& Flushing (proposed in\cite{bailey_border})
&
No stalling, no complex address generation logic, no extra temporary row buffers
&
Extra multiplexers, extra temporary buffers within pixel cache.
\\
\bottomrule
\end{tabularx}
\end{table}

These techniques can be implemented in hardware in a number of ways, as summarized in Table~\ref{table3}.
Further details can be found in \cite{bailey_border}.
Both direct window input and cached priming have the undesired side-effect of extra stalling cycles when processing border pixels.
This can be troublesome in a real-time streaming system, since the regular data movement is disturbed and this significantly complicates datapath control.
The overlapped priming and flushing schemes (naive and the scheme proposed in \cite{bailey_border}) have do not have this effect and hence preserve the regular dataflow.
Both of these techniques require extra multiplexers that ensure replacement values are present in the window for computation.
The naive overlapped priming and flushing scheme requires extra temporary row buffers besides the additional temporary registers within the window pixel cache, resulting in additional use of on-chip memory.

\section{Synthesis Results}

This section explores the effect of both DSP block layout and the border management scheme on implementation area and performance in FPGAs.
All results are post place-and-route.

It is possible to implement spatial filters with no border management.
This, however, means that the output image will be smaller than the input image, since there is no way to compute pixel values for the outermost pixels.
This does, however, simplify control and allow implementation of a transpose form filter.
First we compare such implementations with no border management schemes, including the transpose form, and the direct form with different types of adder trees.
The transpose form does not require a separate adder tree as the individual products are added locally, and this can be achieved using the adder inside the DSP block.

Both the overlapped priming and flushing scheme presented in \cite{bailey_border} and the na\"{i}ve scheme were implemented using Verilog and synthesized and placed and routed on a XC6VLX240T device using Xilinx ISE 14.2. The image specifications were set as follows: bits per pixel ($B=8$), window size ($w=7$), image size $640\times 480$. A filter with general-purpose multipliers such as this can be used for $7\times 7$, $5 \times 5$, and $3 \times 3$ filtering by setting the necessary coefficients to zero.

\begin{table}[h!]
  \caption{Direct and transposed implementation summary with $Image Width (IW)=640$ and $w=7$}
   \label{dir_vs_trnspsd}
 \renewcommand{\arraystretch}{1.2}
   \centering
 \begin{center}
 \begin{tabular}{@{}lrrrcrrr@{}}
 \toprule
 \multirow{2}{*}{Module} & \multicolumn{3}{@{}c@{}}{Direct} & \phantom{x} & \multicolumn{3}{@{}c@{}}{Transposed} \\
 \cmidrule{2-4}\cmidrule{6-8}
  & Reg. & LUT & DSP & & Reg. & LUT & DSP   \\
 \midrule
 Coef. File & 735 & 92  & - & & 735 & 160  & -   \\
 Cntrl. Unit & 47 & 42  & - & & 40 & 31  & -   \\
 Pxl. Cache & 384 &  968 & - & & \multirow{2}{*}{$\Big\}$918} &  \multirow{2}{*}{3668} & \multirow{2}{*}{49} \\
 Fltr. Func. & 3444 & 3101 & 49 & &  &    &  \\
 \midrule
 $F_{\mathit{max}}$ (MHz) & \multicolumn{3}{c}{422} & & \multicolumn{3}{c}{400}  \\

 Latency (Cyc) & \multicolumn{3}{c}{3856} & &  \multicolumn{3}{c}{3850} \\
 \bottomrule
 \end{tabular}
 \end{center}
 \end{table}

\subsection{Direct vs. Transposed Implementations}
\label{subsec:dir-transposed}

Table~\ref{dir_vs_trnspsd} compares the resource utilisation, maximum operating frequency, and latency for both direct and transposed form implementations.
Border exception is not taken into consideration for these designs.
Note that the transpose form combines the pixel cache and filter function into a single module and so separate results are not available for these.
Both implementations are comparable in terms of latency and operating frequency.
However, the direct form consumes significantly more resources than the transposed form.
This is because the transpose form does not require a tree adder, instead using an adder chain that can be packed into the same DSP blocks that implement the individual multipliers.
Note that extra logic is still required for the transpose form image filter, as the dedicated cascade wires that enable transpose form one dimensional filters to be implemented with no extra logic are not suited to the two-dimensional image processing arrangement.
For the direct form, the adder tree must be implemented in the FPGA fabric. The latency can be computed as the following: $IW \times (w-1) + w + \mathit{computation\ latency}$. Note that this latency is not the same as in Table~\ref{table:latency_general} as it, similar to the transposed form, discards border pixels.

While it is clear that the transpose form is favourable, its implementation becomes problematic with border handling as pixel values within the window are always already accumulated with other pixels.
This problem has been discussed in \cite{Benkrid}, where shift registers are employed to overcome this.
However, that approach splits the multiplication and addition apart, which makes packing them into a single DSP block infeasible.
That then means the adder chain must be implemented separately, and the resource consumption becomes similar to that of the direct form.
As a result, the remainder of our experiments we consider the direct form implementation with different tree adder structures and border management techniques.

\subsection{Tree Adder Structure in Direct Implementation}
\label{sub:tree-adder}

Table \ref{dir_no_brdr} compares the 3 adder structure implementations with the direct form filter implementation: \textbf{DSP}, \textbf{LOG}, and \textbf{DSPCOMP}. Border exception is not taken into consideration here.

All three implementations offer high throughput, though the DSP implementation of the adder chain offers around 8\% improved frequency.
In terms of area, the \emph{LOG} Implementation does not use any DSPs for the adder tree. The \emph{DSP} implementation uses 36 DSP blocks, while the \emph{DSPCOMP} implementation uses 20 DSP blocks for $w=7$.
This is in line with the numbers described in Table \ref{table:dsp_block_general}.
Interestingly, while the DSP-based approaches might be expected to use fewer logic resources, the requirement for wide pipeline registers means that in fact they use more registers than the LOG implementation, while the \emph{DSP} implementation does use fewer LUTs.

 \begin{table}[h!]
  \caption{Direct form implementation with different adder tree designs, for $Image Width (IW)=640$ and $w=7$ and no border policy.}
   \label{dir_no_brdr}
 \begin{minipage}[b]{1.0\linewidth}\centering
 \renewcommand{\arraystretch}{1.2}
   \centering
 \begin{center}
 \begin{tabular}{@{}llllll@{}}
 \toprule
 \multirow{4}{*}{Modules}  & \multicolumn{4}{@{}c@{}}{Direct} \\
 \cmidrule{2-5}
 & & \multirow{2}{*}{\begin{tabular}{@{}c@{}} Slice\\[-2pt] Reg. \end{tabular}} & \multicolumn{1}{@{}c@{}}{\multirow{2}{*}{LUTs}} & \multirow{2}{*}{DSPs}  \\
 \\
 \midrule
 Coef. File && 735 & 124  & --    \\
 Cntrl. Unit && 47 & 45 & --    \\
 Pxl. Cache && 384 &  968 & --  \\
 \cmidrule{2-5}
 Fltr. Func. & \multirow{3}{*}{\begin{tabular}{@{}l@{}}
 DSP\\ LOG \\ DSPCOMP \end{tabular}} & 4038 & 2833 & 85
 \\
 &&3444 & 3101 & 49
 \\
 &&5199 & 3453 & 69
 \\
 \midrule
 \multirow{3}{*}{\begin{tabular}{@{}l@{}}$F_{max}$ \\ (MHz) \end{tabular}} &\multirow{3}{*}{\begin{tabular}{@{}l@{}}
 DSP\\ LOG \\ DSPCOMP \end{tabular}}
 &\multicolumn{3}{c}{462}   \\
 &&\multicolumn{3}{c}{422}   \\
 &&\multicolumn{3}{c}{426}   \\

 \cmidrule{1-5}
 \multirow{3}{*}{\begin{tabular}{@{}l@{}}Latency \\ (Cycles) \end{tabular}} &\multirow{3}{*}{\begin{tabular}{@{}l@{}}
 DSP\\ LOG \\ DSPCOMP \end{tabular}}
 &\multicolumn{3}{c}{3874}   \\
 &&\multicolumn{3}{c}{3856}   \\
 &&\multicolumn{3}{c}{3880}   \\
 \bottomrule
 \end{tabular}
 \end{center}
 \end{minipage}
 \end{table}

\subsection{Border Management Overhead}
 \label{subsec:brdr-management}

Having established baseline performance and area results, we now consider border management.
We investigate the effect of adding the overlapped priming and flushing scheme presented in \cite{bailey_border}.
Table \ref{table:brdr_policy_ovrhd_res} shows the logic utilisation for these schemes.
It is clear that adding border management increases complexity.

 \begin{table}[h!]
  \small
 \caption{Logic resources utilisation of pixel cache for different border schemes, for $R=100$, $B=8$, $w=7$}
  \label{table:brdr_policy_ovrhd_res}
  \begin{minipage}[b]{1.0\linewidth}\centering
  \renewcommand{\arraystretch}{1.2}
    \centering
  \begin{center}
\begin{tabular*}{0.9\linewidth}{p{1.5in}ccc}
\toprule
 & \multirow{2}{*}{\begin{tabular}{c} Slice \\ Reg.\end{tabular}} & \multirow{2}{*}{LUTs} \\

 & \multicolumn{1}{l}{} &  \\
\midrule
No Border Policy & 392 & 200 \\
\midrule
Na\"{i}ve overlapped Scheme & 600 & 572 \\
\midrule
\cite{bailey_border}'s Scheme & 552 & 484 \\
\bottomrule
\end{tabular*}

\end{center}
\end{minipage}
 \end{table}

\subsection{Direct Implementation with Border Management}
\label{subsec:dir-brdr}

Table \ref{table:dir_bailey}  shows the direct form filter implementation with the border management technique presented in \cite{bailey_border}.

\begin{table}[t!]
 \caption{Direct implementation using three different adder trees with $Image Width (IW)=640$ and $w=7$ and with border policy from \cite{bailey_border}.}
  \label{table:dir_bailey}
\begin{minipage}[b]{1.0\linewidth}\centering
\renewcommand{\arraystretch}{1.2}
  \centering
\begin{center}
\begin{tabular}{@{}lllll@{}}
\toprule
\multirow{4}{*}{Modules}  & \multicolumn{3}{@{}c@{}}{Direct} \\
\cmidrule{2-5}
& & \multirow{2}{*}{\begin{tabular}{@{}c@{}} Slice\\[-2pt] Reg. \end{tabular}} & \multirow{2}{*}{LUTs} & \multirow{2}{*}{DSPs}  \\
 &&  &  &      \\
\midrule
Coef. File && 735 & 140  & --    \\
Cntrl. Unit && 53 & 69  & --    \\
Pxl. Cache && 552 & 1252 & --  \\
\cmidrule{2-2}
Fil. Func. & \multirow{3}{*}{\begin{tabular}{@{}l@{}}
DSP\\ LOG \\ DSPCOMP \end{tabular}} & 4038 & 1871 & 85
\\
&&3453 & 2160 & 49
\\
&&5199 & 2685 & 69
\\
\midrule
\multirow{3}{*}{\begin{tabular}{@{}l@{}}$F_{max}$ \\ (MHz) \end{tabular}} &\multirow{3}{*}{\begin{tabular}{@{}l@{}}
DSP\\ LOG \\ DSPCOMP \end{tabular}}
&\multicolumn{3}{c}{405}   \\
&&\multicolumn{3}{c}{403}   \\
&&\multicolumn{3}{c}{401}   \\

\cmidrule{1-5}
\multirow{3}{*}{\begin{tabular}{@{}l@{}}Latency \\ (Cycles) \end{tabular}} &\multirow{3}{*}{\begin{tabular}{@{}l@{}}
DSP\\ LOG \\ DSPCOMP \end{tabular}}
&\multicolumn{3}{c}{1951}   \\
&&\multicolumn{3}{c}{1933}   \\
&&\multicolumn{3}{c}{1957}   \\
\bottomrule
\end{tabular}
\end{center}
\end{minipage}
\end{table}

We have seen that a direct form implementation of two-dimensional image filters is amenable to the integration of border management techniques.
We have also seen that different adder tree implementations can be optimised to provide comparable performance, though a logic-only implementation is the most efficient in terms of area.
By taking the low-level structure of the DSP block into account, we have been able to design an architecture that is capable of pixel processing rates close to the theoretical maximum supported by the architecture.

\subsection{Comparison with Vivado HLS Filters}
\label{subsec:hls_filter}

We have also implemented a spatial filter using the Vivado HLS tool to provide a reference for achievable frequenc, this time targetting the Xilinx Zynq architecture.
Both filters were parameterized to process 8-bit ($1920\times 1080$) images with a $7\times7$ window. The Vivado HLS filter only supports fixed coefficients, which allows for architecture optimisation at compile time, but limits runtime flexibility. As a result, the Vivado HLS implementation uses fewer multipliers than the 49 normally required for a $7\times 7$ window (since certain coefficients can be implemented more efficiently in logic)
Table~\ref{table:hls_comparison} shows a comparison of the Vivado HLS implementation with the (\textbf{LOG}) implementation for this paper on a Xilinx Zynq. Vivado HLS infers 7 Block RAMs, likely for data buffers. Both implementations can accommodate a video stream of ($1920\times 1080$) pixels at 60 frames per second, but our implementation has a maximum pixel rate $1.7x$ faster than the HLS implementation, while also providing runtime coefficient flexibility.

 \begin{table}[h!]
 \caption{Resources and timing comparison with Vivado HLS generated filter for ($1920\times 1080$) images on Xilinx Zynq.}
  \label{table:hls_comparison}
\begin{minipage}[b]{1.0\linewidth}\centering
\renewcommand{\arraystretch}{1.2}
  \centering
\begin{center}
\begin{tabular*}{\linewidth}{p{0.8in}cc}

\toprule

 & Vivado HLS Design & \textbf{LOG} Design \\

\midrule
Slice Reg.  & 4129 & 4675\\
\midrule
LUTs  & 3552 & 5287 \\
\midrule
DSPs  & 3 & 49\\
\midrule
RAMs & 7 & - \\
\midrule
$F_{max}$ (MHz) & 214 & 369 \\
\bottomrule
\end{tabular*}

\end{center}
\end{minipage}
\end{table}

\section{Conclusion}
This paper has compared two-dimensional image filter designs including transpose and direct form approaches.
For direct form, different adder tree designs were compared.
Considered use of the DSP blocks found in modern FPGAs results in an architecture that can achieve throughput close to the theoretical maximum for the architecture.
Through detailed architecture-aware design, these filters can process a $640\times 480$ video stream at over 1300 frames per second, or Full HD ($1920\times 1080$) video at over 190 frames per second while taking care of border pixels.
The presented designs are released for use by the community and can be found at \url{https://github.com/ash-aldujaili/spatial-filter-hdl}.

\bibliographystyle{IEEEtran}
\bibliography{references}
%



\end{document}